\documentclass[letter]{aa} 

\usepackage{graphicx}
\usepackage{amsmath}
\usepackage{graphics, color}
\usepackage{txfonts}
\usepackage{natbib}
\begin{document}

   \title{On the radiation driven alignment of dust grains:\\
  Detection of the polarization hole in a starless core\thanks{Based on data acquired with the Atacama 
   Pathfinder Experiment (APEX) and the 1.6 meter telescope at Observatorio do Pico dos Dias (LNA/MCTI).}}



   \author{F. O. Alves
          \inst{1,}\thanks{Present address: Max-Planck-Institut f\"ur extraterrestrische Physik, Giessenbachstrasse 1,
          85748, Garching, Germany; falves@mpe.mpg.de}
          \and
          P. Frau
          \inst{2,3}
          \and
         J. M. Girart
          \inst{4}   
          \and                
         G. A. P. Franco
          \inst{5}
          \and
          F. P. Santos
          \inst{6}
          \and
          H. Wiesemeyer
          \inst{7}          
          }

   \institute{Argelander-Institut f\"ur Astronomie,
             Auf dem H\"ugel 71, D-53121 Bonn, Germany\\
              \email{falves@astro.uni-bonn.de}
         \and
             Instituto de Ciencia de Materiales de Madrid (CSIC),
              Sor Juana In\'es de la Cruz 3, E-28049 Madrid, Spain
	\and
	   Observatorio Astron\'omico Nacional, 
	   Alfonso XII 3, E-28014 Madrid, Spain
         \and
             Institut de Ci\`encies de l'Espai (CSIC-IEEC),
             Campus UAB, Facultat de Ci\`encies, C5 par 2$^{\rm a}$, E-08193, Bellaterra, Catalunya, Spain
          \and
              Departamento de F\'isica -- ICEx -- UFMG, 
              Caixa Postal 702, 30.123-970 Belo Horizonte, Brazil 
          \and
             Department of Physics and Astronomy, Northwestern University, 
             2145 Sheridan Road, Evanston, IL 60208, USA
 	\and
             Max-Planck-Institut f\"ur Radioastronomie, 
             Auf dem H\"ugel 69, D-53121 Bonn, Germany
            }

   \date{Received month day, year; accepted month day, year}

 
  \abstract
     {}
     {We aim to investigate the polarization properties of a starless core
     in a very early evolutionary stage. Linear polarization data 
     reveal the properties of the dust grains in the distinct phases
     of the interstellar medium. Our goal is to investigate how the
     polarization degree and angle correlate with the cloud and core gas.}
   {We use optical, near infrared and submillimeter
   polarization observations toward the starless
   object Pipe-109 in the Pipe nebula. Our data cover a physical 
   scale range of 0.08 to 0.4 pc, comprising the dense gas, envelope and the surrounding cloud.}
   {The cloud polarization is well traced by the optical data. The
    near infrared polarization is 
   produced by a mixed population of grains from the core border and the 
   cloud gas.  The optical and near infrared polarization toward the cloud reach the maximum possible value and 
   saturate with respect to the visual extinction. 
   The core polarization is predominantly traced by the submillimeter data
   and have a steep decrease with respect to the visual extinction. Modeling
   of the submillimeter polarization indicates a magnetic field main direction 
   projected onto the plane-of-sky and loss of grain alignment for densities higher than $6\times10^4$ cm$^{-3}$
   (or $A_V > 30$~mag).}
    {Pipe-109 is immersed in a magnetized medium, with a very ordered magnetic
    field. The absence of internal source of radiation significantly affects the polarization
   efficiencies in the core, creating a polarization hole at the center of the starless core. This result supports the
   theory of dust grain alignment via radiative torques}

   \keywords{Stars: formation -- 
   	      ISM: magnetic fields --
                Techniques: polarimetric --
                ISM: individual objects: Pipe-109
               }
               
\authorrunning{Alves et al.}
   \maketitle
%

\section{Introduction}

Polarimetry is a unique technique to study the physical properties of dust grains. Magnetic fields 
and the interstellar radiation field are known to interact with dust grains, producing polarized light 
\citep{Hall49,Hiltner49}.  The properties of the dust cloud, such as temperature and 
volume density, strongly affect the polarization power of grains \citep[e.g.,][]{Serk75,Goodman95}. 
In addition, polarimetry is a useful tool to characterize the magnetic field morphology. For example, 
linear polarization observations across a wide range of frequencies reveal the morphologic changes 
on the magnetic field topology from molecular clouds down to cloud cores 
\citep[e.g.,][]{Alves11b,Alves12}. 

This paper is focused on the very early stages of star-formation in a highly magnetized environment. 
Our goal is to investigate the polarization properties of a textbook pre-stellar object: Pipe-109\footnote{Here 
we follow the core numbering of \citet{Rathborne08}. It is also known as dark globule FeSt 1-457
in the catalogue of \citet{Feitzinger84} and is comprised by field 40 in the catalog of \citet{Franco10}.}. 
This source is a starless core located in a very pristine cloud, the Pipe nebula
\citep{JAlves07,  Alves08, Forbrich09, Franco10, Frau10, Frau12}. Pipe-109 has a mass of 
$\sim 4$ M$_{\odot}$ and the chemistry of a typical evolved pre-stellar core \citep{Frau12L}.
 
Polarization observations of starless cores are scarce. Only a handful of relevant dust polarization 
data of this class of objects have been published \citep{Ward00,Crutcher04,Nutter04,Ward09}. 
In this paper, we show for the first time combined optical, near infrared (near-IR) and submillimeter 
(submm) polarimetric data of a starless core.  In all cases, we assume that the
polarization is produced by aspherical dust grains aligned perpendicular to the ambient magnetic 
field  \citep[for a detailed review of the mechanisms of grain alignment, refer to][]{Lazarian07}. 
Optical and near-IR polarization are produced by differential absorption of the background 
radiation by the aligned grains. On the other hand, submm polarization is produced by the 
thermal continuum emission from the aligned dust grains. Therefore, the plane-of-sky component of 
the magnetic field is parallel to optical/near-IR polarization maps and perpendicular to 
submm polarization maps. Our submm data  represent one of the few strong detections of dust 
continuum polarized flux toward starless cores. 

\section{Observations}

\subsection{Optical/near infrared observations}

The optical data were obtained in $R$-band (6474 \AA) and a detailed description of the 
observations and data reduction is found in \citet{Franco10}.
The near-IR ($H$-band) linear polarization observations were conducted at the  
Observat\'orio do Pico dos Dias/Laborat\'orio Nacional de Astrof\'isica 
(OPD/LNA, Brazil) using IAGPOL, the IAG imaging polarimeter, mounted on the 1.6\,m 
telescope. For a full description of the polarimeter see \citet{Magalhaes96}. The images were 
gathered using the CamIV infrared camera, which is based on a
HAWAII detector of 1024 $\times$ 1024 pixels and 18.5\,$\mu$m/pixel that yields a
plate scale of $0\farcs25$/pixel.  Sixty dithered images, following a five dots pattern 
(12 $\times$ 5 positions), were obtained for each of eight waveplate positions separated by 
$22\fdg5$. The exposure time for each image was of 10 seconds, totalizing a combined 
exposure time of 600 seconds per waveplate position. The reference direction of the polarizer 
was determined by observing polarized standard stars. Polarization degree and position 
angle were obtained for 700 stars, almost 45\% of which have $P/\sigma_P \ge 10$.

\subsection{Submm observations}

The submm polarization observations were performed with the PolKa continuum polarimeter 
at the APEX 12-m telescope. PolKa is operated with the LABOCA bolometer at 345 GHz and it 
uses a rotating half-wave plate as polarization modulator \citep[for a detailed description of the 
instrument, see][]{Siringo04,Siringo12,Wie14}. The main beam of the telescope at this wavelength 
is $\sim 20^{\prime\prime}$. There was a total of 600 scans of 2.5 minutes each, leading to a 25 
hours on-source time.  Each scan consists of four subscans with a spiral stroke pattern, each 
centered on a different position. The mean zenith opacity was 0.2 (i.e. good weather conditions).
The level of instrumental polarization of our data is about $0.10 \pm 0.04\%$ toward the peak of 
emission and it was determined by observing Uranus. The data discussed below is corrected by 
instrumental polarization. The final submm map has an {\it rms} noise of $\sim 5$ mJy/beam.

\section{Results}
\label{sec:res}

    \begin{figure}[ht!]
    	\centering
	\includegraphics[width=\columnwidth]{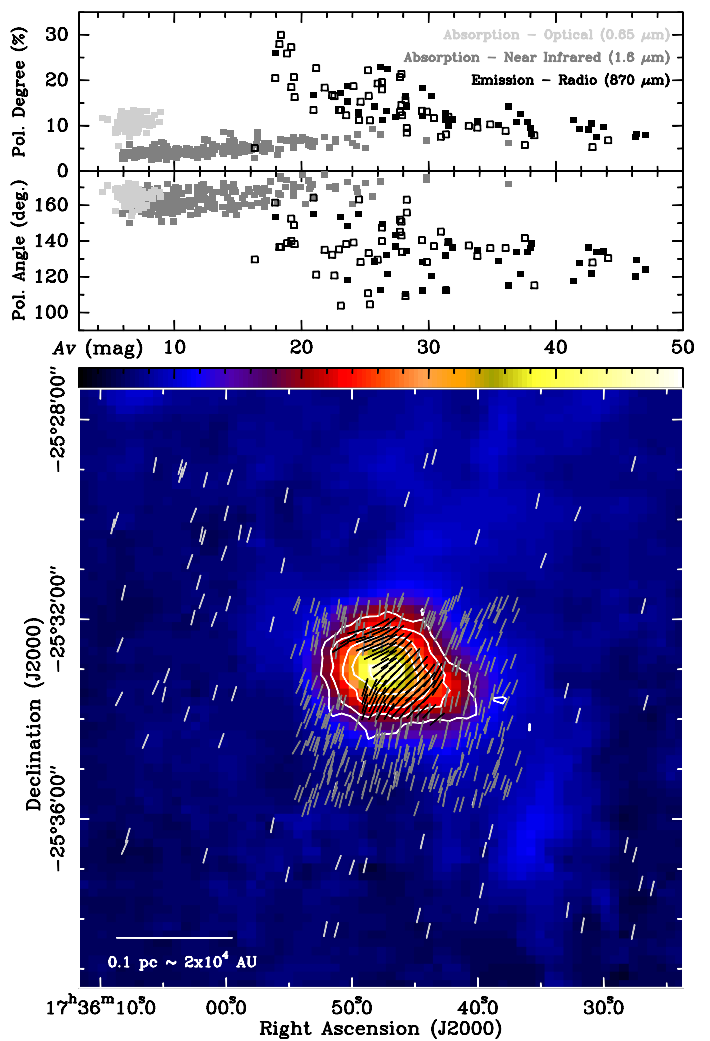}
	\caption{Multi-wavelength polarimetry toward Pipe-109: in the lower panel, 
	optical (light grey), near-IR (grey) 
     	and submm (black) polarization vectors are plotted over the visual extinction map of \citet{Roman10}. 
	The vectors 	are scaled to the same size. The white contours are the 870~$\mu$m dust continuum 
	emission detected by PolKa. 	The contours are 10, 20, 30, 40 and 50 times the {\it rms} of the 
	map ($\sim 5$ mJy beam$^{-1}$). The submm vectors are rotated by 90$\degr$ in order to show 
	the magnetic field direction. The dependence of the polarization degree and polarization angle 
	of the three wavelengths with respect to the visual extinction is shown at the upper 
      	panels. Light grey and grey squares refer to the optical and near-IR data, respectively. 
	Filled black squares and open squares refer to the submm data at 
	$P/\sigma_P > 3$ and $2$, respectively.} \label{fig:pol}
   \end{figure}

Our results are exhibited in Figure \ref{fig:pol}.  We have used the visual extinction map of 
\citet{Roman10}, which traces accurately the gas column density of the core.  It ranges from 
a few magnitudes toward the diffuse gas around the core up to $\sim 45$~mag 
at the core center. The upper panels of Figure \ref{fig:pol} show how the polarization degree 
and position angle of the magnetic field (PA) for the three bands vary with the visual extinction 
($A_V$). The subsequent discussion considers only polarization data whose signal-to-noise ratio 
$P/\sigma_{P}$ is better than 10 (for optical and near-IR) and 3 (for the submm).  

The submm emission has peak intensity of $224\pm5$ mJy/beam and traces only the region with 
$A_V$ higher than 10~magnitudes (mag). 
Note that a combined Bonnor-Ebert fit to the visual extinction and 1.2~mm continuum maps 
shows that the Pipe-109 starts to contribute to the
visual extinction at  $\simeq 9.5$~mag (Frau et al., {\it in prep.}), with lower values being associated only to the 
cloud. Thus, the submm map is sensitive mainly to the starless core.  The submm dust emission is 
optically thin:  assuming a dust 
temperature of  $T_{\mathrm{dust}} \sim 9.5$~K \citep{Rathborne08} we obtain a dust opacity of 
$\sim 10^{-3}$.  The weighted mean polarization is $\sim$ 11\%, with maximum polarized flux of 
$\sim 13$ mJy/beam. Because of the higher sensitivity needed, the submm polarization is sensitive 
to $A_V \ga 18$~mag. The top panel of Fig.~\ref{fig:pol} shows  that the polarization degree
clearly decreases with the visual extinction: it goes from  20-30\% at  $A_V \sim 18$~mag down to
8\% at the highest visual extinction, 45~mag.  The magnetic field lines are relatively uniform with a weighted 
mean position angle of 130$\degr$ (counted from North to East direction) with a standard deviation 
of $\sim 12\degr$. The magnetic field shows more structure at lower extinctions, as inferred by the
increase on PA dispersion for $A_V< 30$~mag ($A_V$--PA panel of Fig.~\ref{fig:pol}). 

The near-IR data is sensitive to $A_V$ in the 6--36~mag, whereas the optical data 
traces the lowest visual extinction of the observed field, 4--9~mag.   The weighted 
mean polarization degree for the optical and near-IR is 11\% and 4\%, respectively. The near-IR
polarization degree exhibits a smooth but clear increase at increasing visual extinction, especially
at $A_V \ga 10$~mag, reaching polarization degrees close to 10\% at $A_V \simeq 25$~mag.
Both data trace a very uniform magnetic field pattern, specially at low visual extinction 
\citep[previously reported by][]{Franco10}.  Thus, the optical and near-IR data have a comparable 
weighted PA of $\sim 
168\degr$ and $163\degr$, respectively.  Their standard deviations are only 4$\degr$ and  5$\degr$. The 
remarkable uniformity of the optical and near-IR data can be also seen in the $A_V$--PA plot. 

A noticeable result of Fig~\ref{fig:pol} is the clear distinction between the two polarization regimes, 
absorption and emission. First, the mean magnetic field direction shown in the submm map, tracing 
the core's field, is clearly bended  by $\simeq 35\degr$ with respect to the optical and near-IR data, 
which mostly trace the cloud magnetic field. The near-IR and submm polarization data overlap in 
the region with a visual extinction in the 18--30~mag range. In this range, the observed discrepancy 
between the magnetic field directions of the near-IR and submm data is clearly seen in the
northern and southwestern section of the core. However, they appear to be consistent in the 
southeastern and eastern parts of the core.  
Second, the polarization degree dependence with respect to the visual extinction shows an
opposite behavior for the polarization data seen in absorption and emission. 
While the  polarization degree is almost constant with respect to the visual extinction for the 
optical and near-IR data for  $A_V \la 10$~mag and then  smoothly increases gradually up to $A_V \sim 20$~mag, the submm polarization degree decreases dramatically, especially in the near-IR and submm overlapping range. 

\section{The cloud and core polarization regimes}   

\begin{figure} 
   \centering
   \includegraphics[width=\hsize]{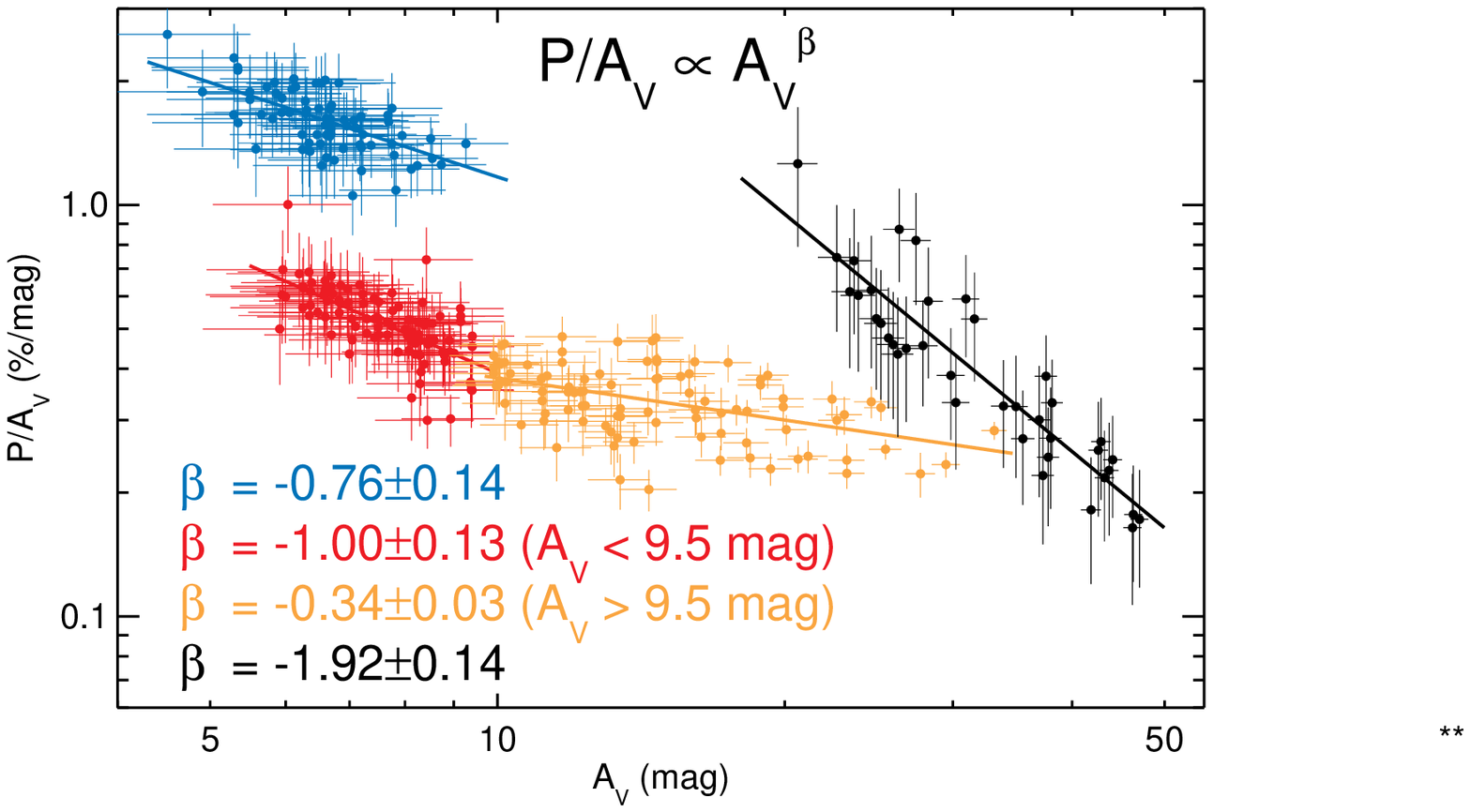}
     \caption{Polarization efficiency ($P/A_V$) as a function of the visual extinction $A_V$ for the
     three wavelengths. The blue dots are the optical data, red dots are near-IR for 
     $A_V < 9.5$~ mag,  orange dots are near-IR data for $A_V > 9.5$ mag and black dots are the 
     submm data. The lines show the linear fit in the log-log basis for the aforementioned four different data. 
      }
         \label{fig:ext}
   \end{figure}

To better understand the observed polarization properties  we have computed the
polarization efficiency for the three bands with respect to the interstellar extinction 
map. The polarization efficiency, defined as the ratio between polarization degree and the visual 
extinction ($P/A_V$), determines the polarizing power of the dust grains. This is shown in Figure \ref{fig:ext}.    
We adopted a conservative uncertainty of 1~mag for the extinction map, which is higher than the value
reported by \citet{Roman10}. Fig. \ref{fig:ext} shows that 
the polarization efficiency decreases in all three wavelengths but with different slopes.  Thus, we 
performed a linear fit on a log-log basis for the different wavelengths. The optical polarization, produced 
uniquely by the cloud, shows a polarization efficiency slope of  $-0.76\pm0.14$ with respect to the 
extinction.  This is consistent with a regime where the polarization is almost saturated 
\citep[e.g.][]{Arce98}.  The near-IR data have a clear breakpoint at 
$\simeq$10~mag, which is noticeably the point where the core starts to contribute. Below this value, 
where the near-IR arises only from the cloud, the slope is closer to full saturation than the optical 
data and is consistent with previous observations in other dark clouds \citep{Goodman95}. However, 
above $\sim$ 10~mag the slope is shallower, $-0.34 \pm 0.03$.  This implies that $P \propto A_V^{0.66}$. 
This power law is similar to the one found in the dense regions of Perseus and Taurus 
\citep{Whittet08, Chapman11}, and is the predicted one for the case of grain alignment mechanism via 
radiative torques  \citep{Dolginov76,Draine96b,Draine97,Lazarian07b}.  
The change of the slope suggests that  for $A_V > 10$~mag the near-IR  polarization is 
produced by a mixed population of grains from both the cloud and  the starless core and  that the core 
grains  are more efficiently aligned than the cloud grains. Indeed, \citet{Goodman95} proposed that the dust population 
in cores and their envelopes have a distinct size distribution, asymmetry level and chemical composition. 

The submm polarization efficiency, which arises uniquely from the core, falls with 
$\sim -1.9$ with respect to $A_V$, which suggests strong depolarization.
Since the submm polarization shows an ordered field, the decrease in polarization 
efficiency can not be fully ascribed to turbulence in the cloud. Molecular emission from dense gas tracers 
at the center of the core shows linewidths of $0.2 - 0.4$ km s$^{-1}$, implying subsonic non-thermal gas 
motions \citep{Frau10,Frau12}.  Since the radiative torques mechanism requires a radiation field, the submm depolarization is thus 
likely produced by a decrease on the grain alignment efficiency due to the lack of internal (no protostar) and external (absorbed UV photons) radiation.  
This is tested in the following section.

A puzzling issue is that in the overlap region between the near-IR and submm polarization there is 
a discrepancy between the two regimes in the polarization efficiency behavior 
and in the derived magnetic field direction (toward the north). 
The distinct slopes for the overlapping extinction range
strongly suggests that there are two populations of dust grains within the core: one population is more sensitive to the near-IR background radiation than the other. They possibly respond to the external radiation in a different 
way, leading to distinct grain alignment regimes (as we see in Fig. \ref{fig:ext}). In addition, the discontinuity in magnetic field direction may be due to physical asymmetries of the core, 
where the two populations are unevenly distributed over the outer layers of the core 
and therefore exposed differently to external ultraviolet (UV) radiation. Preferential grain alignment due to 
inhomogeneities in the cloud was previously reported by other groups \citep[e.g.,][]{Whittet08,Andersson11} and
could be the case here.

\section{The core's depolarization critical density}
     
We made use of a theoretical model to understand the origin of the submm dust polarization and its 
relationship to the source density. Pipe-109 is likely to be on the verge of collapse, and hence, a good 
approximation is the initial pivotal stage used in core collapse simulations. The works by \citet{Li96} and 
\citet{Allen03a,Allen03b} match the observational properties of Pipe-109. They describe the initial stage 
of the magnetic field for an isolated,  marginally supercritical, slightly oblate spheroid under ideal-MHD 
conditions. 

\begin{figure}
   \centering
   \includegraphics[width=\hsize]{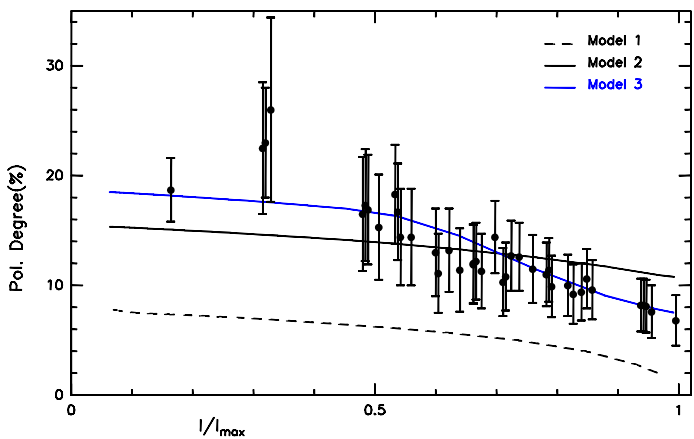}
      \caption{Submillimeter polarization degree as a function of the submillimeter intensity normalized to its peak. 
      The dots represent the PolKa data. Models 1 and 2: magnetized core with constant polarization efficiency 
      ($\alpha = 0.15$) and with a magnetic field direction with respect to the line of sight of $45\degr$ 
      and $90\degr$ (edge-on), respectively. Model 1 clearly underestimates the polarization degree. Model 3: 
      edge-on magnetic field with $\alpha = 0.18$ for densities $n_{\rm (H_2)} < 6 \times 10^4$~cm$^{-3}$.}
   \label{fig:effal}
   \end{figure}

To generate synthetic maps comparable to the data, we used the DustPol module \citep{Padovani12} 
to compute the polarization emission expected for the core under the previous assumptions. Initially, 
we tested the case of a constant polarizing efficiency parameter $\alpha$, which includes the absorption 
cross section and the alignment efficiency \citep{Fiege00}. By adopting the commonly used value of 
$\alpha =0.15$ \citep[e.g.,][]{Frau11, Padovani12} we find that the observed high polarization degree 
can only be achieved if the magnetic field direction is almost on the plane of the sky (Fig.~\ref{fig:effal}). 
However, in this case the model fails to reproduce both the low ($I/I_{\mathrm{max}}<0.4$) and high 
($I/I_{\mathrm{max}}>0.8$) values of the normalized intensity. This suggests a higher/lower polarization 
efficiency at low/high normalized intensity values.  To test this, we assumed a scenario with a constant 
but higher polarization efficiency,  $\alpha =0.18$, up to a certain density, $n_{\rm depol}$ , becoming 
zero at higher densities. The observational data can be well reproduced for a depolarization density of   
$n_{\rm depol} = 6 \times 10^4$~cm$^{-3}$ (model 3 in Fig.~\ref{fig:effal}), which corresponds to 
a visual extinction limit of 30~mag. In brief, the Pipe-109 
submm polarization suggests that the core's center is unpolarized. These results supports 
the  radiative torques theory of grain alignment which requires a source of photons to help the alignment 
\citep{Lazarian07b}.
   
\section{Conclusions}       

Our multi-wavelength data represent the most complete polarimetric observations ever performed toward a starless core. This investigation shows that the magnetic 
field has a reasonable uniform morphology from physical scales of 0.4 pc down to 0.08 pc. However,
a discontinuity between the optical/near-IR and the submm polarization can be seen in the mean field direction.

The absorption and emission polarization regimes are well correlated to the cloud and core components, 
respectively. While the optical and near-IR polarization do not vary significantly with the diffuse/low 
extinction gas ($A_V < 9$ magnitudes), the submm polarization falls dramatically toward the peak of extinction. 
The near-IR polarization is produced by both the cloud and the core, resulting in two distinct regimes
of polarization efficiencies of the dust grains. The near-IR polarization of the core overlaps in extinction with the 
submm polarization, but the cause for the distinct polarization efficiency regimes between them is still unclear. Further work should be done in order to understand this issue.
The submm regime shows a steep decrease in polarization efficiency. The lack of an internal source of 
radiation results in a loss of grain alignment with the magnetic field. This result is consistent with the theory
of dust grain alignment via radiative torques.

The submm polarization is well reproduced by a magnetized, isolated core model whose magnetic field
is projected onto the plane of sky. The observed depolarization occurs for volume densities
higher than $6 \times 10^4$ cm$^{-3}$ (or $A_V > 30$~mag).


\begin{acknowledgements}
      The authors would like to thank  the APEX staff for the pool observations and the 
      OPD staff for helping with the optical and near-IR observations. The authors would also like to 
      thank Carlos  Rom\'an-Z\'u\~niga for providing the visual extinction map. 
      FOA acknowledges financial support from the BMBF Verbundforschung 05A11PD3 project.
      GAPF and FPS acknowledge support from the Brazilian agencies CNPq, CAPES and FAPEMIG.
      FPS is supported by the CAPES grant 2397/13-7.
      PF is supported by the Spanish CONSOLIDER project CSD2009-00038. PF and JMG are supported 
      by the Spanish MINECO AYA2011-30228-C03-02, and Catalan AGAUR 2009SGR1172 grants.
\end{acknowledgements}


\end{document}